\pdfoutput=1

\documentclass[aps,pra,10pt,amsmath,twocolumn,floatfix,superscriptaddress,longbibliography,eqsecnum]{revtex4-1}
\usepackage[dvipsnames]{xcolor}
\usepackage[colorlinks=true,bookmarks=false,linkcolor=NavyBlue,urlcolor=NavyBlue,citecolor=NavyBlue,breaklinks]{hyperref}
\usepackage{amsmath,amsfonts,amssymb,amsthm}
\usepackage{mathtools}
\usepackage{tabularx}
\usepackage{graphicx}
\usepackage[italicdiff]{physics}
\usepackage{times}
\usepackage{enumitem}
\usepackage{setspace}

\newcommand{\intall}{\int_{-\infty}^{\infty}}
\newcommand{\avg}[1]{\langle#1\rangle}
\newcommand{\Avg}[1]{\left\langle#1\right\rangle}
\newcommand{\bs}[1]{\boldsymbol{#1}}
\newcommand{\floor}[1]{\left\lfloor#1\right\rfloor}

\newcommand{\bk}[1]{\left(#1\right)}
\newcommand{\Bk}[1]{\left[#1\right]}
\newcommand{\BK}[1]{\left\{#1\right\}}

\newcommand{\mc}[1]{\mathcal #1}
\newcommand{\ms}[1]{\mathsf #1}
\renewcommand{\norm}[1]{\Vert #1 \Vert}

\renewcommand{\exp}{\operatorname{exp}}

\DeclareMathOperator{\spn}{span}
\DeclareMathOperator{\cspn}{\overline{span}}

\DeclareMathOperator{\expect}{\mathbb E}

\newcommand{\cgraphic}[2]{\centerline{\includegraphics[width=#1\textwidth]{#2}}}

\begin{document}

\title{Quantum limit to subdiffraction incoherent optical imaging.
III. Numerical analysis}

\author{Xiao-Jie Tan}
\affiliation{Department of Electrical and Computer Engineering,
  National University of Singapore, 4 Engineering Drive 3, Singapore
  117583}

\author{Mankei Tsang}
\email{mankei@nus.edu.sg}
\homepage{https://blog.nus.edu.sg/mankei/}
\affiliation{Department of Electrical and Computer Engineering,
  National University of Singapore, 4 Engineering Drive 3, Singapore
  117583}

\affiliation{Department of Physics, National University of Singapore,
  2 Science Drive 3, Singapore 117551}

\date{\today}

%\pacs{42.50.Wk, 03.65.Ta, 42.65.Yj}

\begin{abstract}
  To investigate the fundamental limit to far-field incoherent
  imaging, the prequels to this work [M.~Tsang, Phys.~Rev.~A
  \textbf{99}, 012305 (2019); \textbf{104}, 052411 (2021)] have
  studied a quantum lower bound on the error of estimating an object
  moment and proved a scaling law for the bound with respect to the
  object size. As the scaling law was proved only in the asymptotic
  limit of vanishing object size, this work performs a numerical
  analysis of the quantum bound to verify that the law works well for
  nonzero object sizes in reality. We also use the numerical bounds to
  study the optimality of a measurement called spatial-mode
  demultiplexing or SPADE, showing that SPADE not only follows the
  scaling but is also numerically close to being optimal, at least for
  low-order moments.
\end{abstract}

\maketitle
\section{Introduction}
The quantum nature of light imposes fundamental limits to the
precision of optical measurements for information processing.  To
derive the most fundamental quantum limits, Helstrom invented a theory
of quantum detection and estimation---also called quantum metrology in
modern terminology---and first studied its implications for optics
\cite{helstrom}. More recently, this line of research has yielded a
pleasant surprise: the quantum limits to incoherent imaging turn out
to be far less stringent than previously thought, but still achievable
by physical measurements \cite{tnl,review_cp}. These results have
potential impact on both optical astronomy and fluorescence microscopy
\cite{review_cp}.

Beyond the assumption of two point sources in the earlier literature,
much recent effort has been devoted to the case of multiple point
sources. In particular, the prequels to this work
\cite{qlmoment_pra,qlmoment_pra2} (henceforth called paper I and paper
II, respectively) study a quantum Cram\'er-Rao lower bound on the
error of estimating an object moment; see also Ref.~\cite{zhou19}. As
those works aim to derive the bound for an unknown distribution of
many incoherent point sources, the dimension of the parameter space
(or equivalently the number of scalar parameters) may be high or even
infinite, making the analysis formidable and an exact expression for
the bound elusive. In classical statistics, a problem or any concept
associated with an infinite-dimensional parameter space is called
semiparametric.  Even in the classical case, semiparametric problems
are difficult and require advanced mathematics \cite{bickel}.

Despite the daunting semiparametric setting of the moment estimation
problem, progress has been made---Paper I proposes a scaling law for
the quantum bound with respect to the object size, while paper II
proves the scaling law rigorously by using the quantum semiparametric
theory developed in Ref.~\cite{semi_prx}. The measurement method of
spatial-mode demultiplexing (SPADE) can achieve the same scaling
\cite{qlmoment_pra2,spade_njp,spade_pra,spade_prr}, but since the
scaling law is proved only in the asymptotic limit of vanishing object
size, it remains an open question how well the law holds for nonzero
object sizes in reality. Another open question is the gap between the
quantum limit and the SPADE error. This question is significant
because a small gap would imply that SPADE is almost optimal, while a
large gap would leave open the possibility that another measurement
can significantly improve moment estimation but is yet to be
discovered. The moment estimation problem also has significant
implications for more general image reconstruction problems
\cite{superosc_ieee} and random displacement models
\cite{noise_spec_pra}, so it is worthwhile to devise a method that can
compute quantum bounds more precisely for future studies along these
lines.

To verify the scaling law and to investigate the gap between the
quantum limit and the SPADE error, this work proposes a numerical
method to compute a quantum Cram\'er-Rao bound for moment estimation.
The numerical results demonstrate that the scaling law works well for
nonzero object sizes, not just in asymptotics, and the SPADE errors
are close to the quantum bounds, at least for low moment orders. For
higher-order moments, the results are unfortunately less conclusive.

The proposed method is nontrivial because the bound needs to be valid
for an infinite-dimensional parameter space, yet computable with
finite resources. To achieve this goal, we use the concept of
parametric submodels, which is established in classical semiparametric
statistics \cite{bickel} and generalized for the quantum case by
Ref.~\cite{semi_prx}. Paper II has previously used the submodel
concept to prove the scaling law analytically; here the approach is
extended to create a numerical method.

Apart from the aforementioned papers, the closest prior work may be
Ref.~\cite{bisketzi19} by Bisketzi and coworkers, which numerically
computes quantum bounds for the localization of multiple point
sources. The most important difference between their method and ours
is that, whereas their method works only for a finite number of point
sources, our method can produce bounds for extended objects modeled by
continuous intensity functions. Reference~\cite{bisketzi19} also
mentions moment estimation only in passing. Other outstanding prior
works that deal with extended objects include
Refs.~\cite{dutton19,prasad20,prasad23}, although those works assume
objects with known shapes and low-dimensional parameter spaces (i.e.,
few scalar parameters) and have not explicitly shown that their bounds
are also valid for semiparametric problems.

\section{Theory}

\subsection{Quantum optics}
Following paper I and II, we consider the far-field imaging of
one-dimensional spatially incoherent optical sources. A weak-source
approximation \cite{stellar,tnl,review_cp} results in the density
operator
\begin{align}
\rho &= (1-\epsilon)\tau_0 + \epsilon \tau
\end{align}
for each temporal optical mode, where $\tau_0$ is the vacuum state,
$\tau$ is the one-photon state on a Hilbert space $\mc H$ that models
the spatial degrees of freedom, and $\epsilon \ll 1$ is the average
photon number per temporal mode, With $M$ temporal modes, the density
operator is assumed to be
\begin{align}
\omega &= \rho^{\otimes M},
\label{omega}
\end{align}
and the average photon number in all modes is then
\begin{align}
N \equiv M\epsilon.
\end{align}
For one-dimensional spatially incoherent sources, $\tau$ can be
expressed as
\begin{align}
\tau &= \intall F(x) e^{-i\hat k x}\ket{\psi}\bra{\psi} e^{i\hat k x} dx,
\end{align}
where $F$ is the nonnegative density function for the sources with the
normalization $\intall F(x) dx = 1$, $\ket{\psi} \in \mc H$ with
$\braket{\psi}{\psi} = 1$ models the point-spread function of the
imaging system, and $\hat k$ is a self-adjoint momentum operator on
$\mc H$ such that $\exp(-i\hat k x)$ models the photon displacement.
To be specific, let $\ket{k}$ be the Dirac eigenket of $\hat k$ such
that $\hat k\ket{k} = k \ket{k}$ and $\braket{k}{k'} = \delta(k-k')$.
Then the optical transfer function is given by $\braket{k}{\psi}$ and
the point-spread function for the optical field is given by
\begin{align}
\psi(x) &\equiv \frac{1}{\sqrt{2\pi}} \intall e^{ikx} \braket{k}{\psi} dk.
\end{align}

To obtain tighter quantum bounds, we generalize paper I and II
slightly by assuming that both $\epsilon$ and $F$ may depend on
unknown parameters.  Then it is convenient to use the theory of
Poisson states recently proposed in Ref.~\cite{poisson_quantum}, which
simplifies quantum information calculations when the Poisson limit
\begin{align}
\epsilon &\to 0,
&
M &\to \infty,
&
N \textrm{ fixed}
\label{poisson}
\end{align}
is taken. A Poisson state is completely specified by the so-called
intensity operator
\begin{align}
\Gamma &\equiv N \tau,
\end{align}
and quantum bounds in the Poisson limit can be computed by considering
only $\Gamma$ rather than $\epsilon$ and $\tau$. For the imaging
problem, we can write
\begin{align}
\Gamma(G) &= \intall G(x) e^{-i\hat k x}\ket{\psi}\bra{\psi} e^{i\hat k x} dx,
\end{align}
where $G$ is an unnormalized object intensity function given by
\begin{align}
G(x) &\equiv N F(x).
\end{align}

\subsection{Quantum Cram\'er-Rao bounds}
To express the quantum bounds succinctly, we first establish some
basic notations. Given an intensity operator $\Gamma$, define a
weighted inner product between two Hermitian operators $h$ and $g$ as
\begin{align}
\Avg{h,g}_{\Gamma} &\equiv \trace (h\circ g)\Gamma,
\end{align}
where
\begin{align}
h\circ g &\equiv \frac{1}{2}\bk{hg + gh}
\end{align}
denotes the Jordan product. The corresponding norm is defined as
\begin{align}
\norm{h}_\Gamma &\equiv \sqrt{\avg{h,h}_\Gamma}.
\end{align}
%A real Hilbert space $L_2(\Gamma)$ can then be constructed by
%completing the space of bounded self-adjoint operators with respect to
%the norm \cite{holevo11}.
Similarly, define the inner product
\begin{align}
\Avg{h,g}_G &\equiv \intall G(x) h(x) g(x) dx 
\end{align}
between real c-number functions $h,g$ with respect to an object
intensity $G$. The associated norm is then
$\norm{h}_G \equiv \sqrt{\avg{h,h}_G}$.
% and the associated real Hilbert space is denoted as $L_2(G)$.

Now let the set of all possible object intensity functions
in the estimation problem be
\begin{align}
\bs G &\equiv \BK{G: G(x) \ge 0, \intall G(x) dx  < \infty},
\label{G}
\end{align}
such that the set of intensity operators is
\begin{align}
\Gamma(\bs G) \equiv \BK{\Gamma(G): G \in \bs G}.
\end{align}
Let $\beta:\bs G \to \mathbb R$ be a real scalar parameter of interest
and $G_0 \in \bs G$ be the true object intensity. To define a quantum
Cram\'er-Rao bound for this semiparametric estimation problem,
consider a one-dimensional parametric submodel expressed as
\begin{align}
\bs\gamma = \BK{\gamma_\theta = \Gamma(G_\theta): \theta \in \Theta
\subseteq \mathbb R} \subset \Gamma(\bs G).
\end{align}
$\bs\gamma$ is required to contain the true $\Gamma(G_0)$, and a
parametrization has been chosen to give $\gamma_0 = \Gamma(G_0)$
without loss of generality.  The Hermitian score operator $S$ of the
submodel is defined in terms of the intensity operator $\gamma_\theta$
by
\begin{align}
\partial\gamma_\theta &= \gamma_0 \circ S.
&
\partial(\cdot) &\equiv \left.\pdv{(\cdot)}{\theta}\right|_{\theta=0}.
\end{align}
The Helstrom bound $\ms H_{\bs\gamma}$ for the submodel of the
Poisson states is then \cite{poisson_quantum}
\begin{align}
\ms H_{\bs\gamma} &\equiv \frac{[\partial\beta(G_\theta)]^2}
{\norm{S}_{\gamma_0}^2}.
\label{helstrom_sub}
\end{align}
$\ms H_{\bs\gamma}$ is a lower bound on the mean-square error $\ms E$
of any unbiased estimator and any measurement for the semiparametric
problem; the precise definition of $\ms E$ is in
Appendix~\ref{app_ghb}. Appendix~\ref{app_ghb} also shows that the
generalized Helstrom bound $\tilde{\ms H}$ for the semiparametric
problem \cite{semi_prx} can be expressed as the supremum of all the
submodel bounds, viz.,
\begin{align}
\tilde{\ms H} &= \sup_{\bs\gamma} \ms H_{\bs\gamma},
\label{ghb}
\end{align}
such that, for any submodel $\bs\gamma$,
\begin{align}
\ms E &\ge \tilde{\ms H} \ge \ms H_{\bs\gamma}.
\end{align}

\subsection{\label{sec_moment}Quantum bounds for moment estimation}
Suppose that the point-spread function $\psi(x)$ has a normalized
width $1$, $G(x)$ is centered at $x = 0$, and the support of $G$ is
infinite with $\sup\{|x|: G(x) > 0\} \equiv \Delta$. The
subdiffraction regime is defined by $\Delta \ll 1$. Let the parameter
of interest be a generalized moment of order
$\mu \in \mathbb N_0 \equiv \{0,1,2,\dots\}$, defined as
\begin{align}
\beta(G) &= \intall G(x) b(x) dx,
&
b(x) &= x^\mu + o(\Delta^\mu),
\label{beta}
\end{align}
where $o[f(\Delta)]$ denotes an order smaller than $f(\Delta)$ as
$\Delta \to 0$.  Other asymptotic notions $O[f(\Delta)]$ [order at
most $f(\Delta)$], $\Omega[f(\Delta)]$ [order at least $f(\Delta)$],
and $\Theta[f(\Delta)]$ [order exactly $f(\Delta)$] may also be used
in this paper. $\tilde{\ms H}$ is hard to compute, so paper II seeks
an unfavorable submodel such that the submodel bound
$\ms H_{\bs\gamma}$ gives the desired scaling law. Assuming
$\beta = \intall F(x) b(x) dx$ in terms of the normalized $F$ and a
submodel with a fixed $N$, paper II proves a submodel bound given by
\begin{align}
\ms H_{\bs\gamma} &=  \frac{\Omega(\Delta^{2\floor{\mu/2}})}{N},
\quad
\mu \ge 1.
\end{align}
For a $\beta$ given by Eqs.~(\ref{beta}) in terms of the unnormalized
$G$, on the other hand, the submodel bound in paper II remains valid
after a simple rescaling and can be expressed as
\begin{align}
\ms H_{\bs\gamma} &=  \Omega(\Delta^{2\floor{\mu/2}}),
\label{submodel_bound}
\end{align}
where we omit the scaling with $N$ for brevity. It can also be shown
that Eq.~(\ref{submodel_bound}) remains valid for $\mu = 0$ by
considering a submodel with $N$ being a function of $\theta$.

To improve upon the submodel used in paper II, here we consider a
$p$-dimensional submodel
\begin{align}
\bs\Upsilon &= \BK{\gamma_\theta = \Gamma(G_\theta): \theta \in
\Theta \subseteq \mathbb R^p} \subset \Gamma(\bs G),
\end{align}
with $\theta = (\theta_0,\theta_1,\dots,\theta_{p-1})$. Its Helstrom
bound is
\begin{align}
\ms H_{\bs\Upsilon} &= u^\top K^{-1} u,
\label{helstrom_p}
\end{align}
where $u$ is the $p\times 1$ column vector
given by
\begin{align}
u_j &\equiv \partial_j\beta(G_\theta),
&
\partial_j(\cdot) &\equiv \left.\pdv{(\cdot)}{\theta_j}\right|_{\theta = 0},
\label{u}
\end{align}
$\top$ is the transpose, and $K$ is the $p\times p$ Helstrom
information matrix given by
\begin{align}
K_{jk} &\equiv \Avg{S_j,S_k}_{\gamma_0},
\label{K}
\\
\partial_j\gamma_\theta &= \gamma_0 \circ S_j.
\label{sld}
\end{align}
$\ms H_{\bs\Upsilon}$ coincides with the maximum Helstrom bound
for all the one-dimensional submodels within $\bs\Upsilon$
\cite{semi_prx,gross20}. It follows that 
\begin{align}
\tilde{\ms H} &\ge \ms H_{\bs\Upsilon}.
\end{align}
Instead of guessing an unfavorable submodel, the computation of
$\ms H_{\bs\Upsilon}$ allows us to obtain directly the bound for the
least favorable submodel within $\bs\Upsilon$.

The parametrization of the object intensity $G_\theta(x)$ affects the
bound only through the true intensity $G_0(x)$ and the derivatives
$\partial_jG_\theta(x)$ at $\theta = 0$, since
\begin{align}
u_j &= \intall \partial_jG_\theta(x) b(x) dx,
\\
\partial_j\gamma_\theta &= 
\intall \partial_jG_\theta(x)
e^{-i\hat k x}\ket{\psi}\bra{\psi} e^{i\hat k x} dx.
\label{dgamma}
\end{align}
Taking advantage of this fact, the submodel approach in paper II picks
a $\partial G_\theta(x)$ first and then constructs a submodel based on
the derivative.  The derivative chosen there is
\begin{align}
\partial G_\theta(x) &= G_0(x) a_\mu(x),
\label{aj}
\end{align}
where $a_\mu(x)$ is an orthonormal polynomial with respect to $G_0$
in the sense of
\begin{align}
a_j(x) &= \sum_{k=0}^j A_{jk} x^k,
&
\Avg{a_j,a_k}_{G_0} &= \delta_{jk},
\end{align}
and $\mu$ in Eq.~(\ref{aj}) is chosen to match the order of the moment
parameter of interest. We are guaranteed to do no worse than the
submodel bound in paper II if we assume a $p$-dimensional submodel
that obeys
\begin{align}
\partial_j G_\theta(x) &= G_0(x) a_j(x),
&
j = 0,\dots,\mu,\dots,p-1,
\label{aj2}
\end{align}
such that it includes the one-dimensional submodel used in paper II.
An example parametrization that satisfies all the desired properties
is
\begin{align}
G_\theta(x) &= \BK{1 + \tanh\Bk{\sum_{j=0}^{p-1}\theta_j a_j(x)}} G_0(x).
\end{align}
The $\tanh$ function ensures that each derivative of $G_\theta(x)$
agrees with Eq.~(\ref{aj2}) and also $G_\theta(x)$ is a valid
intensity function as per Eq.~(\ref{G}).

To demonstrate the relations among all the aforementioned bounds more
explicitly, we write the $u$ in the submodel Helstrom bound
$\ms H_{\bs\Upsilon}$ given by Eq.~(\ref{helstrom_p}) as
\begin{align}
u_j &= \intall G_0(x) a_j(x) b(x) dx  = \Avg{a_j,b}_{G_0},
\label{uj}
\end{align}
and write the relation between $a_j$ and the score operator $S_j$
through Eqs.~(\ref{sld}), (\ref{dgamma}), and (\ref{aj2}) in terms of
a linear map $\Gamma_*$ as
\begin{align}
S_j &= \Gamma_* a_j.
\end{align}
$\Gamma_*$ is called a pushforward \cite{qlmoment_pra2} or a
generalized conditional expectation \cite{hayashi,gce_pra,gce2}.  As
shown in Appendix~\ref{app_helstrom_max}, $\ms H_{\bs\Upsilon}$ can be
expressed as
\begin{align}
\ms H_{\bs\Upsilon} &= \max_{s \in \spn\{a_0,\dots,a_{p-1}\}}
\frac{\avg{s,b}_{G_0}^2}{\norm{\Gamma_* s}_{\Gamma(G_0)}^2},
\label{helstrom_alt}
\end{align}
where $\spn$ denotes the linear span. This expression makes it clear
that $\ms H_{\bs\Upsilon}$ for the $p$-dimensional submodel cannot be
lower than the submodel bound $\ms H_{\bs\gamma}$ considered in paper
II, given by
\begin{align}
\ms H_{\bs\gamma} &= \frac{\avg{a_\mu,b}_{G_0}^2}
{\norm{\Gamma_*a_\mu}_{\Gamma(G_0)}^2}.
\label{helstrom_II}
\end{align}
On the other hand, following Eq.~(\ref{ghb}), the generalized Helstrom
bound can be expressed as
\begin{align}
\tilde{\ms H} &= \sup_{s: \norm{\Gamma_* s}_{\Gamma(G_0)}<\infty}
\frac{\avg{s,b}_{G_0}^2}{\norm{\Gamma_* s}_{\Gamma(G_0)}^2}.
\label{helstrom_gen}
\end{align}
Then
\begin{align}
\ms E &\ge \tilde{\ms H}  \ge \ms H_{\bs\Upsilon} \ge \ms H_{\bs\gamma}
\end{align}
holds for Eqs.~(\ref{helstrom_alt})--(\ref{helstrom_gen}). We are,
however, unable to prove $\ms H_{\bs\Upsilon} \to \tilde{\ms H}$ in
the $p\to \infty$ limit, because it is unknown whether
$\spn\{a_j:j \in \mathbb N_0\}$ is dense in the space
$\{s:\norm{\Gamma_* s}_{\Gamma(G_0)}<\infty\}$ needed for
Eq.~(\ref{helstrom_gen}).

\section{Numerical analysis}
\subsection{\label{sec_meth}Method}
Assuming a parameter of interest given by Eqs.~(\ref{beta}), the
submodel Helstrom bound $\ms H_{\bs\Upsilon}$ can be computed from
Eqs.~(\ref{helstrom_p}), (\ref{K}), (\ref{sld}), (\ref{dgamma}),
(\ref{aj2}), and (\ref{uj}).  The coefficients $A_{jk}$ of the
orthonormal polynomials $\{a_j\}$ can be computed by standard linear
algebra; see, for example, Appendix~A in paper II. Note that
\begin{align}
\Avg{a_j,x^\mu}_{G_0} = 0 \textrm{ if }j > \mu.
\end{align}
If the parameter of interest $\beta$ is a simple moment with
$b(x) \propto x^\mu$, then $u_j = 0$ for $j > \mu$, meaning that
$\beta$ has no local sensitivity to such a $\theta_j$, and it suffices
to compute the submodel bound with $j \le p-1 = \mu$. For a
generalized moment with order $\mu$, the diminished sensitivity of
$\beta$ to the higher-order parameters should similarly make our bound
somewhat insensitive to a restriction to the parameter dimension $p$
as long as $p -1 \ge \mu$.

We numerically solve the Lyapunov equation in Eq.~(\ref{sld}) and
compute the Helstrom information matrix $K$ in Eq.~(\ref{K}) by
working in the point-spread-function-adapted (PAD) basis
\cite{spade_pra,rehacek17}
\begin{align}
\BK{\ket{\psi_n} = (-i)^n \tilde a_n(\hat k)\ket{\psi}: n = 0,1,\dots,q-1},
\end{align}
where $\{\tilde a_n\}$ are the orthonormal polynomials with respect to
$|\braket{k}{\psi}|^2$. We choose the PAD basis because the matrix
entries of $\gamma_0$ and $\partial_j\gamma_\theta$ in this basis
decay quickly for subdiffraction objects, thus helping us reduce the
errors due to the necessary truncation of the matrices in the
numerics. We use MATLAB on a personal computer to perform the
computation and its \texttt{lyap} function to solve the Lyapunov
equation.

To model the performance of SPADE, we follow paper II and take $b(x)$
for an even moment order to be
\begin{align}
b(x) &= \frac{1}{D_n^2} \abs{C_n(x)}^2,
\label{b_even}
\\
C_n(x) &\equiv \bra{\psi_n} e^{-i\hat k x}\ket{\psi}
= D_{n} x^n + o(\Delta^{n}).
\end{align}
$D_n$ can be shown to be real and positive.  $\beta$ is then a
generalized moment of order $\mu = 2n$ as per Eqs.~(\ref{beta}) and
can be estimated by a measurement of the PAD mode $\ket{\psi_n}$.  Let
the detected photon number in the mode be $m_n$, which is Poisson
under the limit given by Eqs.~(\ref{poisson}). Under the Poisson
limit, the expected value of $m_n$ is
\begin{align}
\expect(m_n) &=  \bra{\psi_n}\Gamma(G_0)\ket{\psi_n}
\\
&= \intall G_0(X) |C_n(x)|^2 dx = D_n^2 \beta.
\end{align}
An unbiased estimator is then
\begin{align}
\check\beta &= \frac{m_n}{D_n^2},
\label{beta_even_est}
\end{align}
and its mean-square error is
\begin{align}
\ms E &= \frac{\expect(m_n)}{D_n^4} = \frac{\beta}{D_n^2} = O(\Delta^\mu).
\label{spade_even}
\end{align}
For an odd moment order, we take
\begin{align}
b(x) &= \frac{1}{D_{n}D_{n+1}}\Re[C_n(x)C_{n+1}^*(x)],
\label{b_odd}
\end{align}
such that $\beta$ is a generalized moment of order $\mu = 2n+1$ and
can be estimated by a measurement of the interferometric PAD modes
\cite{spade_njp,spade_pra}
\begin{align}
\ket{\psi_n^+} &\equiv \frac{1}{\sqrt{2}}\bk{\ket{\psi_n}+\ket{\psi_{n+1}}},
\\
\ket{\psi_n^-} &\equiv \frac{1}{\sqrt{2}}\bk{\ket{\psi_n}-\ket{\psi_{n+1}}}.
\end{align}
Let the detected photon number in the plus
mode be $m_{n}^+$ and that in the minus mode be $m_{n}^-$ and assume
again the Poisson limit.  Their expected values are respectively given
by
\begin{align}
\expect\bk{m_{n}^+} &= \bra{\psi_n^+}\Gamma(G_0)\ket{\psi_n^+}
\\
&= \frac{1}{2} \intall G_0(x) \abs{C_n(x) + C_{n+1}(x)}^2 dx,
\\
\expect\bk{m_{n}^-} &= \bra{\psi_n^-}\Gamma(G_0)\ket{\psi_n^-}
\\
&= \frac{1}{2} \intall G_0(x) \abs{C_n(x) - C_{n+1}(x)}^2 dx.
\end{align}
An unbiased estimator is then
\begin{align}
\check\beta &= \frac{m_{n}^+ - m_{n}^-}{2 D_n D_{n+1}},
\label{beta_odd_est}
\end{align}
and its error is
\begin{align}
  \ms E &= \frac{\expect(m_{n}^+) + \expect(m_n^-)}{4 D_n^2 D_{n+1}^2} 
= O(\Delta^{\mu-1}).
\label{spade_odd}
\end{align}

Given the quantum bound $\ms H_{\bs\Upsilon}$ for a certain submodel
$\{G_\theta\}$ and the SPADE error $\ms E$ at a certain $G_0$, the
bound for the submodel $\{c G_\theta\}$ and the error at $c G_0$,
where $c$ is a known constant, are scaled simply by $c$. Since our
goal here is to study the ratio between $\ms H_{\bs\Upsilon}$ and
$\ms E$ and their scalings with $\Delta$, there is no loss of
generality if we assume that $G_0$ happens to be normalized.  Assume,
furthermore, that $G_0$ is the rectangle function given by
\begin{align}
G_0(x) &= 
\begin{cases}
1/(2\Delta), & |x| \le \Delta,\\
0, & \textrm{otherwise}.
\end{cases}
\end{align}
$\ms H_{\bs\Upsilon}$ and $\ms E$ can now be computed numerically.

We note that the superiority of SPADE over direct imaging for moment
estimation has been shown many times before
\cite{qlmoment_pra2,spade_njp,spade_pra,spade_prr,superosc_ieee}, so
here we do not study direct imaging and focus only on a comparison
between SPADE and the quantum limit.

\subsection{Results}
\begin{figure}[htbp!]
\cgraphic{0.48}{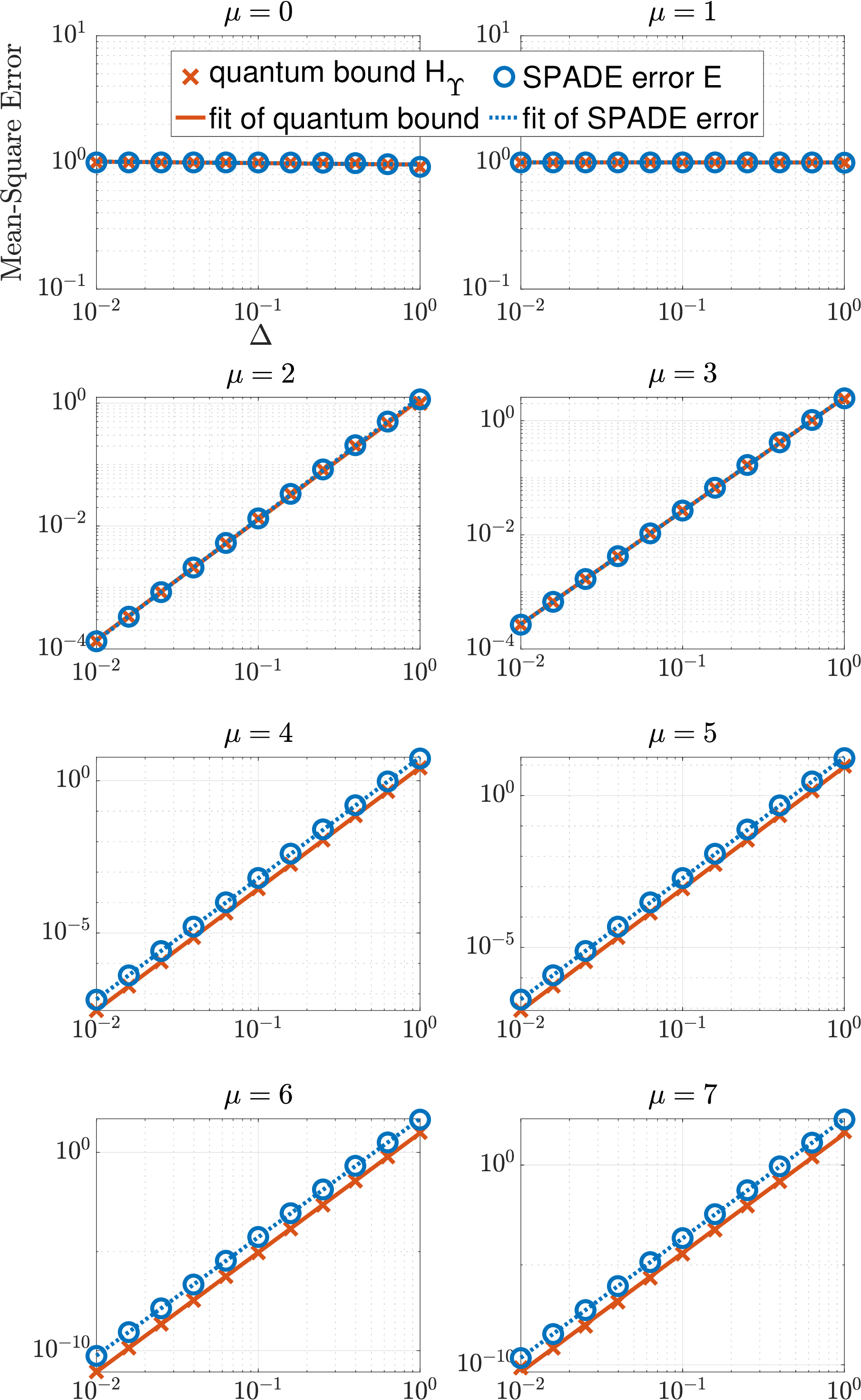}
\caption{\label{gauss}Plots of the numerically computed Helstrom bound
  $\ms H_{\bs\Upsilon}$ (crosses) and the SPADE mean-square error
  $\ms E$ (circles) versus the object size $\Delta$ in log-log scale,
  assuming the Gaussian transfer function given by
  Eq.~(\ref{gauss_transfer}). $\mu$ is the order of the moment
  parameter of interest. The lines are straight-line fits as per
  Eqs.~(\ref{H_fit}) and (\ref{E_fit}) and the fitting coefficients
  are given in Table~\ref{gauss_fit}. Both axes are dimensionless.}
\end{figure}

We first assume the Gaussian optical transfer function
\begin{align}
\braket{k}{\psi} &= 
\bk{\frac{2}{\pi}}^{1/4}\exp(-k^2),
\label{gauss_transfer}
\end{align}
such that
\begin{align}
\psi(x) &= \frac{1}{(2\pi)^{1/4}}\exp\bk{-\frac{x^2}{4}},
\\
C_n(x) &= D_n x^n \exp\bk{-\frac{x^2}{8}},
\quad
D_n = \frac{1}{2^n\sqrt{n!}}.
\end{align}
With the method described in Sec.~\ref{sec_meth}, we compute the
quanum bound $\ms H_{\bs\Upsilon}$ and the SPADE error $\ms E$ as a
function of the object size $\Delta$. The code assumes a submodel with
$p = 10$ parameter dimensions and truncates the matrices of $\gamma_0$
and $\partial_j\gamma_\theta$ in the PAD basis
$\{\ket{\psi_0},\dots,\ket{\psi_{q-1}}\}$ by taking $q =
6$. Figure~\ref{gauss} plots the results in log-log scale for moment
orders $\mu = 0,\dots,7$, together with the straight-line fits defined
by
\begin{align}
\log_{10} \ms H_{\bs\Upsilon} &\approx \log_{10} \ms H_{\bs\Upsilon}^{(0)} +  
\ms H_{\bs\Upsilon}^{(1)}\log_{10} \Delta,
\label{H_fit}
\\
\log_{10} \ms E &\approx \log_{10} \ms E^{(0)} +  \ms E^{(1)}\log_{10} \Delta,
\label{E_fit}
\end{align}
where the superscript $(0)$ denotes the prefactor and the superscript
$(1)$ denotes the exponent in the scaling with $\Delta$.  We obtain
$\log_{10} \ms H_{\bs\Upsilon}^{(0)}$, $\ms H_{\bs\Upsilon}^{(1)}$,
$\log_{10} \ms E^{(0)}$, and $\ms E^{(1)}$ by passing
$\log_{10}\Delta$ and $\log_{10}\ms H_{\bs\Upsilon}$ or
$\log_{10}\ms E$ to the \texttt{polyfit} function in MATLAB with
degree $1$; the resulting coefficients are reported in
Table~\ref{gauss_fit}. These numbers change by at most $1\%$ even if
we increase the parameter dimension $p$ to $16$ and the Hilbert-space
dimension $q$ to $12$.

\begin{table}[htbp!]
\begin{tabular}{|c|c|c|c|c|c|c|c|c|}
\hline
$\mu$ & 0 & 1 & 2 & 3 & 4 & 5 & 6 & 7 \\
\hline
$\ms H_{\bs\Upsilon}^{(0)}$ & $0.96$ & $1.0$ & $1.2$ & $2.5$ & $2.8$ & $8.9$ & $9.6$ 
& $36$\\
\hline
$\ms E^{(0)}$ & $0.96$ & $1.0$ & $1.2$ & $2.6$ & $5.9$ & $18$ & $50$
 & $200$ \\
\hline
$\ms E^{(0)}/\ms H_{\bs\Upsilon}^{(0)}$ & $1.0$ & $1.0$ & $1.1$ & $1.0$
& $2.1$ & $2.0$ & $5.2$ & $5.6$\\
\hline
$\ms H_{\bs\Upsilon}^{(1)}$ & $0.0$ & $0.0$ & $2.0$ & $2.0$ & $4.0$ & $4.0$ & $6.0$ & $6.0$\\
\hline
$\ms E^{(1)}$ & $0.0$ & $0.0$ & $2.0$ & $2.0$ & $4.0$ & $4.0$ & $6.0$ & $6.0$\\
\hline 
\end{tabular}
\caption{\label{gauss_fit}Coefficients from the straight-line fits of
  the Helstrom bound $\ms H_{\bs\Upsilon}$ and the SPADE error $\ms E$
  versus the object size $\Delta$ in log-log scale, as plotted in
  Fig.~\ref{gauss}, assuming the Gaussian transfer function given by
  Eq.~(\ref{gauss_transfer}). $\mu$ is the order of the moment
  parameter of interest, and the fitting coefficients are defined by
  Eqs.~(\ref{H_fit}) and (\ref{E_fit}). For brevity, the prefactors
  $\ms H_{\bs\Upsilon}^{(0)}$ and $\ms E^{(0)}$ and their ratios
  $\ms E^{(0)}/\ms H_{\bs\Upsilon}^{(0)}$ are rounded to 2 significant
  figures, while the exponents $\ms H_{\bs\Upsilon}^{(1)}$ and
  $\ms E^{(1)}$ are rounded to one decimal place. }
\end{table}

The results in Fig.~\ref{gauss} and Table~\ref{gauss_fit} demonstrate
three significant features. The first is that the exponents
$\ms H_{\bs\Upsilon}^{(1)}$ and $\ms E^{(1)}$ follow closely the
theoretical exponent $2\floor{\mu/2}$ for the considered range of
$\Delta$, hence verifying that the scaling law works well for nonzero
$\Delta$. The second feature is that the SPADE errors are all very
close to the quantum bounds up to moment order $\mu = 3$, as
quantified by the ratios $\ms E^{(0)}/\ms H_{\bs\Upsilon}^{(0)}$ in
Table~\ref{gauss_fit} being close to $1$ and the gaps in
Fig.~\ref{gauss} being close to zero. The small gaps mean that SPADE
is almost quantum-optimal for those moments.  The third feature is
that, for higher moment orders $\mu \ge 4$, the gaps begin to widen,
and the results are less conclusive regarding the optimality of SPADE.

\begin{figure}[htbp!]
\cgraphic{0.48}{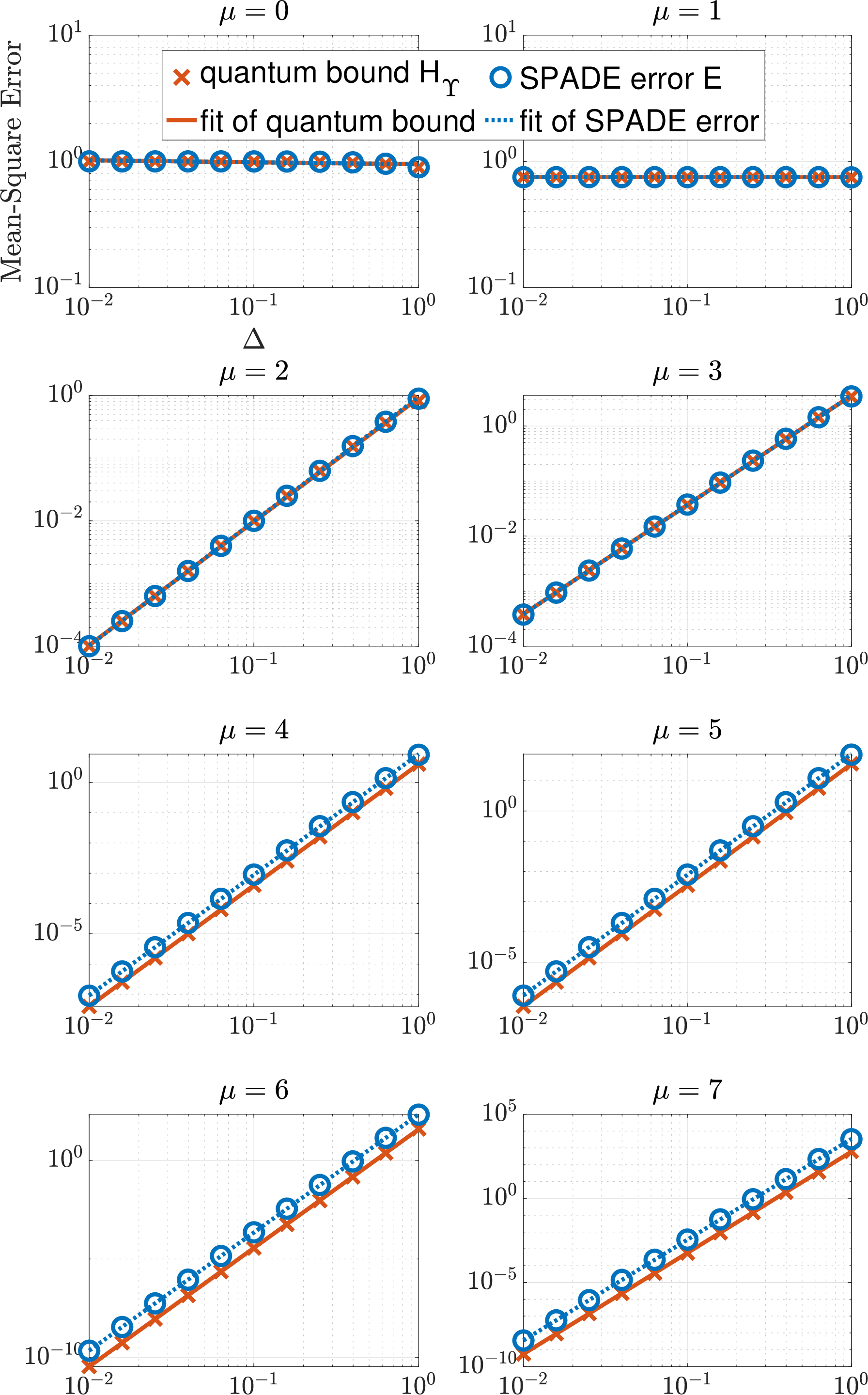}
\caption{\label{bessel}Same as Fig.~\ref{gauss}, except that the
  rectangle transfer function given by Eq.~(\ref{rect_transfer}) is
  assumed.}
\end{figure}

In Fig.~\ref{bessel} and Table~\ref{bessel_fit},
we also report the numerical results for the rectangle
transfer function
\begin{align}
\braket{k}{\psi} &= 
\begin{cases}
1/\sqrt{2}, & |k| \le 1,
\\
0, & \textrm{otherwise},
\end{cases}
\label{rect_transfer}
\end{align}
which leads to
\begin{align}
\psi(x) &= \frac{1}{\sqrt{\pi}} j_0(x) = 
\frac{1}{\sqrt{\pi}}
\begin{cases}
(\sin x)/x, & x \neq 0,
\\
1, & x = 0,
\end{cases}
\\
C_n(x) &= \sqrt{2n+1}j_n(x),
\quad
D_n = \frac{2^n  n!\sqrt{2n+1}}{(2n+1)!},
\end{align}
where $j_n$ is the spherical Bessel function of the first kind
\cite{olver}.  The assumptions and formats of Fig.~\ref{bessel} and
Table~\ref{bessel_fit} otherwise follow those of Fig.~\ref{gauss} and
Table~\ref{gauss_fit}.  These results exhibit the same features
described earlier for the Gaussian transfer function.

\begin{table}[htbp!]
\begin{tabular}{|c|c|c|c|c|c|c|c|c|}
\hline
$\mu$ & 0 & 1 & 2 & 3 & 4 & 5 & 6 & 7 \\
\hline
$\ms H_{\bs\Upsilon}^{(0)}$ & $0.95$ & $0.75$ & $0.91$ & $3.6$ & $4.0$ & $36$ & $38$ & $590$\\
\hline
$\ms E^{(0)}$ & $0.95$ & $0.75$ & $0.94$ & $3.6$ & $8.6$ & $76$ & $220$ & $3400$ \\
\hline
$\ms E^{(0)}/\ms H_{\bs\Upsilon}^{(0)}$ & $1.0$ & $1.0$ & $1.0$ & $1.0$
& $2.1$ & $2.1$ & $5.7$ & $5.8$\\
\hline
$\ms H_{\bs\Upsilon}^{(1)}$ & $0.0$ & $0.0$ & $2.0$ & $2.0$ & $4.0$ & $4.0$ & $6.0$ & $6.0$\\
\hline
$\ms E^{(1)}$ & $0.0$ & $0.0$ & $2.0$ & $2.0$ & $4.0$ & $4.0$ & $6.0$ & $6.0$\\
\hline 
\end{tabular}
\caption{\label{bessel_fit}Same as Table~\ref{gauss_fit}, except that
  the rectangle transfer function given by Eq.~(\ref{rect_transfer})
  is assumed.}
\end{table}

\subsection{\label{sec_limit}Limitations}
We point out a few limitations of our numerical approach.  First, we
are unable to prove that our submodel bound $\ms H_{\bs\Upsilon}$
approaches the generalized Helstrom bound $\tilde{\ms H}$ even in the
$p \to \infty$ limit, as discussed at the end of
Sec.~\ref{sec_moment}. The potential existence of a higher quantum
bound means that, when the gap between the error and the quantum bound
shown in Fig.~\ref{gauss} or \ref{bessel} is large, we cannot tell
whether it is because the measurement is suboptimal or because the
quantum bound is loose. Second, even $\tilde{\ms H}$ may not be tight
if multiple scalar parameters are to be estimated simultaneously, and
there exist better---albeit harder-to-compute---bounds for that
problem
\cite{holevo11,semi_prx,carollo19,*carollo20,nagaoka89,hayashi99,conlon21}.
Third, the numerical approach must assume a specific true object
intensity $G_0$ and a specific moment order $\mu$ in computing each
bound, so even if the numerical results demonstrate the scaling proved
in paper I and II, the analytic result in paper I and II remains more
general in the sense that the scaling is proved there for a much more
general class of $G_0$ and an arbitrary $\mu$. Fourth, while the
numerical approach can produce the prefactor
$\ms H_{\bs\Upsilon}^{(0)}$ for a given $G_0$ and $\mu$, no rigorous
analytic result is yet available regarding the prefactor as a function
of $\mu$, and the general behavior of the prefactor remains an
interesting open problem.

\section{\label{sec_ext}Potential extensions}
For completeness, here we discuss some straightforward potential
extensions of our theory.  First, it is easy to use our method to
compute the quantum bound for other types of the parameter of interest
$\beta(G)$ by changing $u$ in Eqs.~(\ref{helstrom_p}) and (\ref{u});
$\beta(G)$ can even be nonlinear with respect to $G$. A simple example
is the normalized moment
\begin{align}
\beta(G) &= \frac{1}{N(G)} \intall G(x) b(x) dx,
\label{beta_normalized}
\\
N(G) &\equiv \intall G(x) dx.
\end{align}
To compute $\ms H_{\bs\Upsilon}$ for the normalized moment, one only
needs to change $u_j$ to
\begin{align}
u_j &= \partial_j\beta(G_\theta)
= \Avg{a_j,b}_{F_0} - \Avg{a_j,1}_{F_0} \Avg{b,1}_{F_0},
\label{u_normalized}
\end{align}
where $F_\theta \equiv G_\theta/N(G_\theta)$. Of all the orthonormal
polynomials, only $a_0$ has a nonzero $\avg{a_j,1}_{F_0}$.

Second, if $N$ is known, as assumed in paper I and II, then we can
parametrize the normalized $F$ only and keep $N$ fixed, such that
$G_\theta = N F_\theta$. The normalization implies that all scores
should obey
\begin{align}
\partial_j\Gamma(G_\theta) = \avg{S_j,1}_{\Gamma(G_0)} = \avg{a_j,1}_{G_0} = 0,
\end{align}
meaning that $\theta_0$ and $a_0$ should be excluded from the
submodel.  A valid submodel with $N$ fixed is
\begin{align}
F_\theta(x) &= \frac{\{1+\tanh[\sum_{j=1}^{p} \theta_j a_j(x)]\}
F_0(x)}{\intall (\textrm{numerator}) dx}.
\end{align}
Assuming this submodel and a normalized moment given by
Eq.~(\ref{beta_normalized}), the $u_j$ given by
Eq.~(\ref{u_normalized}) becomes $\avg{a_j,b}_{F_0}$, and we obtain
back the theory assumed in paper II.

Third, when the parameter of interest is a normalized moment given by
Eq.~(\ref{beta_normalized}), the estimator with SPADE should be
changed depending on the situation. If $N$ is known, then one can
simply divide the estimators in Sec.~\ref{sec_meth} by $N$, and the
mean-square errors in Sec.~\ref{sec_meth} are then scaled by $1/N^2$.
If $N$ is unknown, an unbiased estimator can still be constructed if
one can detect the total photon number $L$ in all the modes. Consider
the estimation of an even-order $\beta$ given by Eqs.~(\ref{b_even})
and (\ref{beta_normalized}) and a measurement in the PAD
basis. Conditioned on $L$, the photon numbers $\{m_0,m_1,\dots\}$
detected in the modes now follow the multinomial distribution, and an
unbiased estimator is simply Eq.~(\ref{beta_even_est}) divided by
$L$. The error conditioned on $L$ is now
\begin{align}
\ms E &= 
%\frac{\expect(m_n)}{D_n^4 L^2} \Bk{1-\frac{\expect(m_n)}{L}}= 
\frac{\beta}{D_n^2 L} \bk{1-D_n^2\beta}
= \frac{O(\Delta^\mu)}{L},
\end{align}
which is slightly different from Eq.~(\ref{spade_even}) because of the
multinomial statistics but still has the same optimal scaling with
$\Delta$. The theory for an odd-order $\beta$ is similar.

\section{Conclusion}
Our numerical analysis has verified the object-size scaling law for
the moment estimation error in incoherent optical imaging.  Paper I
and II have proved the law only in the asymptotic limit of vanishing
object size, and our results here show that the law also works well
for nonzero sizes in reality. We have also shown that SPADE is close
to quantum-optimal for moment orders up to $\mu = 3$. For higher
orders, the results are less conclusive, as significant gaps are
observed between the SPADE errors and the quantum bounds. It remains
an open question whether a better measurement can be found or a
tighter bound can be derived.

In terms of future directions, apart from addressing the limitations
discussed in Sec.~\ref{sec_limit} and the potential extensions in
Sec.~\ref{sec_ext}, an important problem is to find the quantum limit
to the reconstruction of the object intensity $G$, rather than just
its moments, as well as the measurement to approach it. A
multidimensional parameter of interest $\beta$ should then be assumed,
and a quantum bound beyond the Helstrom family
\cite{holevo11,semi_prx,carollo19,*carollo20,nagaoka89,hayashi99,conlon21}
may be needed to produce a tighter bound. Finding the measurement to
achieve it will also be a difficult but rewarding problem.

\section*{Acknowledgment}
This research is supported by the National Research Foundation,
Singapore, under its Quantum Engineering Programme (QEP-P7).

\appendix

\section{\label{app_ghb}Generalized Helstrom bound}
Let the set of all possible density operators in a statistical problem
be $\bs\rho = \{\rho_\theta:\theta \in \Theta\}$, and let the true
density operator be $\rho \in \bs\rho$. Let the parameter of interest
be $\beta:\Theta \to \mathbb R$. Consider the density operator
$\omega = \rho^{\otimes M}$ and a measurement modeled by a positive
operator-valued measure $\mc E$, such that the probability of an event
$\Lambda$ upon the measurement is
\begin{align}
P(\Lambda) &= \trace \mc E(\Lambda) \omega = 
\trace \mc E(\Lambda) \rho^{\otimes M}.
\end{align}
Let a measurement outcome be $\lambda$ and an estimator be $\check\beta(\lambda)$.
The mean-square error is
\begin{align}
\ms E &= \int \Bk{\check\beta(\lambda)-\beta}^2 \trace \mc E(d\lambda)\omega.
\end{align}
To derive a lower bound on $\ms E$, consider a one-dimensional parametric submodel
\begin{align}
\bs\sigma &= \BK{\sigma_\theta: \theta \in \Phi \subseteq \mathbb R} \subseteq \bs\rho.
\end{align}
We require $\bs\sigma$ to contain the true $\rho$. Without loss of
generality, assume a parametrization that gives $\sigma_0 =
\rho$. Define the score operator $S$ for the submodel by
\begin{align}
\partial\sigma_\theta &= \sigma_0 \circ S,
\end{align}
let $\{S\}$ be the scores of all submodels with
$\norm{S}_\rho < \infty$, and define the tangent space
$\mc T \equiv \cspn\{S\}$ as the closed linear span. Assume also that
there exists a so-called influence operator $\delta$, defined by the
properties
\begin{align}
\trace \rho \delta &= 0,
&
\avg{\delta,S}_\rho &= \partial\beta
\end{align}
for all submodels. 
For any measurement and any unbiased estimator, the
generalized Helstrom bound $\tilde{\ms H}$ is given by (see
Ref.~\cite{semi_prx} and Lemma~2 in paper II)
\begin{align}
\ms E &\ge \tilde{\ms H} = \frac{1}{M} \norm{\Pi(\delta|\mc T)}^2_\rho
= \sup_{u \in \spn\{S\}}
\frac{\avg{\delta,u}_\rho^2}{M\norm{u}_\rho^2},
\end{align}
where $\Pi(\delta|\mc T)$ is the orthogonal projection of $\delta$
into $\mc T$ and $\spn$ denotes the linear span. We can also make the
benign assumption that any linear combination of scores is the score
of a submodel, or $\spn\{S\} = \{S\}$.  Then $\tilde{\ms H}$ is the
supremum of all the submodel bounds, viz.,
\begin{align}
\tilde{\ms H} &= \sup_{\bs\sigma} \ms H_{\bs\sigma}.
\end{align}
For Poisson states, each submodel bound is given by
Eq.~(\ref{helstrom_sub}) according to Ref.~\cite{poisson_quantum}, and
we can take Eq.~(\ref{ghb}) to be the generalized Helstrom bound in
this paper.

\section{\label{app_helstrom_max}Alternative expression for
the Helstrom bound}
The goal here is to show that Eq.~(\ref{helstrom_alt})
is the same as the Helstrom bound given by Eq.~(\ref{helstrom_p}).
Any $s \in \spn\{a_0,\dots,a_{p-1}\}$ can be expressed as
\begin{align}
s &= \sum_j v_j a_j,
\quad
v \in \mathbb R^p.
\end{align}
Treating $v$ as a column vector and using Eq.~(\ref{uj}), the
right-hand side of Eq.~(\ref{helstrom_alt}) becomes
\begin{align}
\max_{v} \frac{(v^\top u)^2}{v^\top K v}.
\label{helstrom_v}
\end{align}
The Cauchy-Schwartz inequality yields 
\begin{align}
(v^\top u)^2 &= (v^\top K^{1/2} K^{-1/2} u)^2 \le (v^\top K v) (u^\top K^{-1} u),
\end{align}
which makes Eq.~(\ref{helstrom_v}) equal to Eq.~(\ref{helstrom_p}).

\bibliography{research3}

%merlin.mbs apsrev4-1.bst 2010-07-25 4.21a (PWD, AO, DPC) hacked
%Control: key (0)
%Control: author (0) dotless jnrlst
%Control: editor formatted (1) identically to author
%Control: production of article title (0) allowed
%Control: page (1) range
%Control: year (0) verbatim
%Control: production of eprint (0) enabled
\begin{thebibliography}{31}%
\makeatletter
\providecommand \@ifxundefined [1]{%
 \@ifx{#1\undefined}
}%
\providecommand \@ifnum [1]{%
 \ifnum #1\expandafter \@firstoftwo
 \else \expandafter \@secondoftwo
 \fi
}%
\providecommand \@ifx [1]{%
 \ifx #1\expandafter \@firstoftwo
 \else \expandafter \@secondoftwo
 \fi
}%
\providecommand \natexlab [1]{#1}%
\providecommand \enquote  [1]{``#1''}%
\providecommand \bibnamefont  [1]{#1}%
\providecommand \bibfnamefont [1]{#1}%
\providecommand \citenamefont [1]{#1}%
\providecommand \href@noop [0]{\@secondoftwo}%
\providecommand \href [0]{\begingroup \@sanitize@url \@href}%
\providecommand \@href[1]{\@@startlink{#1}\@@href}%
\providecommand \@@href[1]{\endgroup#1\@@endlink}%
\providecommand \@sanitize@url [0]{\catcode `\\12\catcode `\$12\catcode
  `\&12\catcode `\#12\catcode `\^12\catcode `\_12\catcode `\%12\relax}%
\providecommand \@@startlink[1]{}%
\providecommand \@@endlink[0]{}%
\providecommand \url  [0]{\begingroup\@sanitize@url \@url }%
\providecommand \@url [1]{\endgroup\@href {#1}{\urlprefix }}%
\providecommand \urlprefix  [0]{URL }%
\providecommand \Eprint [0]{\href }%
\providecommand \doibase [0]{http://dx.doi.org/}%
\providecommand \selectlanguage [0]{\@gobble}%
\providecommand \bibinfo  [0]{\@secondoftwo}%
\providecommand \bibfield  [0]{\@secondoftwo}%
\providecommand \translation [1]{[#1]}%
\providecommand \BibitemOpen [0]{}%
\providecommand \bibitemStop [0]{}%
\providecommand \bibitemNoStop [0]{.\EOS\space}%
\providecommand \EOS [0]{\spacefactor3000\relax}%
\providecommand \BibitemShut  [1]{\csname bibitem#1\endcsname}%
\let\auto@bib@innerbib\@empty
%</preamble>
\bibitem [{\citenamefont {Helstrom}(1976)}]{helstrom}%
  \BibitemOpen
  \bibfield  {author} {\bibinfo {author} {\bibfnamefont {Carl~W.}\ \bibnamefont
  {Helstrom}},\ }\href
  {http://www.sciencedirect.com/science/bookseries/00765392/123} {\emph
  {\bibinfo {title} {Quantum Detection and Estimation Theory}}}\ (\bibinfo
  {publisher} {Academic Press},\ \bibinfo {address} {New York},\ \bibinfo
  {year} {1976})\BibitemShut {NoStop}%
\bibitem [{\citenamefont {Tsang}\ \emph {et~al.}(2016)\citenamefont {Tsang},
  \citenamefont {Nair},\ and\ \citenamefont {Lu}}]{tnl}%
  \BibitemOpen
  \bibfield  {author} {\bibinfo {author} {\bibfnamefont {Mankei}\ \bibnamefont
  {Tsang}}, \bibinfo {author} {\bibfnamefont {Ranjith}\ \bibnamefont {Nair}}, \
  and\ \bibinfo {author} {\bibfnamefont {Xiao-Ming}\ \bibnamefont {Lu}},\
  }\bibfield  {title} {\enquote {\bibinfo {title} {Quantum theory of
  superresolution for two incoherent optical point sources},}\ }\href {\doibase
  10.1103/PhysRevX.6.031033} {\bibfield  {journal} {\bibinfo  {journal}
  {Physical Review X}\ }\textbf {\bibinfo {volume} {6}},\ \bibinfo {pages}
  {031033} (\bibinfo {year} {2016})}\BibitemShut {NoStop}%
\bibitem [{\citenamefont {Tsang}(2019{\natexlab{a}})}]{review_cp}%
  \BibitemOpen
  \bibfield  {author} {\bibinfo {author} {\bibfnamefont {Mankei}\ \bibnamefont
  {Tsang}},\ }\bibfield  {title} {\enquote {\bibinfo {title} {Resolving
  starlight: a quantum perspective},}\ }\href {\doibase
  10.1080/00107514.2020.1736375} {\bibfield  {journal} {\bibinfo  {journal}
  {Contemporary Physics}\ }\textbf {\bibinfo {volume} {60}},\ \bibinfo {pages}
  {279–298} (\bibinfo {year} {2019}{\natexlab{a}})}\BibitemShut {NoStop}%
\bibitem [{\citenamefont {Tsang}(2019{\natexlab{b}})}]{qlmoment_pra}%
  \BibitemOpen
  \bibfield  {author} {\bibinfo {author} {\bibfnamefont {Mankei}\ \bibnamefont
  {Tsang}},\ }\bibfield  {title} {\enquote {\bibinfo {title} {Quantum limit to
  subdiffraction incoherent optical imaging},}\ }\href {\doibase
  10.1103/PhysRevA.99.012305} {\bibfield  {journal} {\bibinfo  {journal}
  {Physical Review A}\ }\textbf {\bibinfo {volume} {99}},\ \bibinfo {pages}
  {012305} (\bibinfo {year} {2019}{\natexlab{b}})}\BibitemShut {NoStop}%
\bibitem [{\citenamefont {Tsang}(2021{\natexlab{a}})}]{qlmoment_pra2}%
  \BibitemOpen
  \bibfield  {author} {\bibinfo {author} {\bibfnamefont {Mankei}\ \bibnamefont
  {Tsang}},\ }\bibfield  {title} {\enquote {\bibinfo {title} {Quantum limit to
  subdiffraction incoherent optical imaging. {II}. {A} parametric-submodel
  approach},}\ }\href {\doibase 10.1103/PhysRevA.104.052411} {\bibfield
  {journal} {\bibinfo  {journal} {Physical Review A}\ }\textbf {\bibinfo
  {volume} {104}},\ \bibinfo {pages} {052411} (\bibinfo {year}
  {2021}{\natexlab{a}})}\BibitemShut {NoStop}%
\bibitem [{\citenamefont {Zhou}\ and\ \citenamefont {Jiang}(2019)}]{zhou19}%
  \BibitemOpen
  \bibfield  {author} {\bibinfo {author} {\bibfnamefont {Sisi}\ \bibnamefont
  {Zhou}}\ and\ \bibinfo {author} {\bibfnamefont {Liang}\ \bibnamefont
  {Jiang}},\ }\bibfield  {title} {\enquote {\bibinfo {title} {Modern
  description of {Rayleigh}'s criterion},}\ }\href {\doibase
  10.1103/PhysRevA.99.013808} {\bibfield  {journal} {\bibinfo  {journal}
  {Physical Review A}\ }\textbf {\bibinfo {volume} {99}},\ \bibinfo {pages}
  {013808} (\bibinfo {year} {2019})}\BibitemShut {NoStop}%
\bibitem [{\citenamefont {Bickel}\ \emph {et~al.}(1993)\citenamefont {Bickel},
  \citenamefont {Klaassen}, \citenamefont {Ritov},\ and\ \citenamefont
  {Wellner}}]{bickel}%
  \BibitemOpen
  \bibfield  {author} {\bibinfo {author} {\bibfnamefont {Peter~J.}\
  \bibnamefont {Bickel}}, \bibinfo {author} {\bibfnamefont {Chris A.~J.}\
  \bibnamefont {Klaassen}}, \bibinfo {author} {\bibfnamefont {Ya'acov}\
  \bibnamefont {Ritov}}, \ and\ \bibinfo {author} {\bibfnamefont {John~A.}\
  \bibnamefont {Wellner}},\ }\href@noop {} {\emph {\bibinfo {title} {Efficient
  and Adaptive Estimation for Semiparametric Models}}}\ (\bibinfo  {publisher}
  {Springer},\ \bibinfo {address} {New York},\ \bibinfo {year}
  {1993})\BibitemShut {NoStop}%
\bibitem [{\citenamefont {Tsang}\ \emph {et~al.}(2020)\citenamefont {Tsang},
  \citenamefont {Albarelli},\ and\ \citenamefont {Datta}}]{semi_prx}%
  \BibitemOpen
  \bibfield  {author} {\bibinfo {author} {\bibfnamefont {Mankei}\ \bibnamefont
  {Tsang}}, \bibinfo {author} {\bibfnamefont {Francesco}\ \bibnamefont
  {Albarelli}}, \ and\ \bibinfo {author} {\bibfnamefont {Animesh}\ \bibnamefont
  {Datta}},\ }\bibfield  {title} {\enquote {\bibinfo {title} {Quantum
  {Semiparametric} {Estimation}},}\ }\href {\doibase
  10.1103/PhysRevX.10.031023} {\bibfield  {journal} {\bibinfo  {journal}
  {Physical Review X}\ }\textbf {\bibinfo {volume} {10}},\ \bibinfo {pages}
  {031023} (\bibinfo {year} {2020})}\BibitemShut {NoStop}%
\bibitem [{\citenamefont {Tsang}(2017)}]{spade_njp}%
  \BibitemOpen
  \bibfield  {author} {\bibinfo {author} {\bibfnamefont {Mankei}\ \bibnamefont
  {Tsang}},\ }\bibfield  {title} {\enquote {\bibinfo {title} {Subdiffraction
  incoherent optical imaging via spatial-mode demultiplexing},}\ }\href
  {\doibase 10.1088/1367-2630/aa60ee} {\bibfield  {journal} {\bibinfo
  {journal} {New Journal of Physics}\ }\textbf {\bibinfo {volume} {19}},\
  \bibinfo {pages} {023054} (\bibinfo {year} {2017})}\BibitemShut {NoStop}%
\bibitem [{\citenamefont {Tsang}(2018)}]{spade_pra}%
  \BibitemOpen
  \bibfield  {author} {\bibinfo {author} {\bibfnamefont {Mankei}\ \bibnamefont
  {Tsang}},\ }\bibfield  {title} {\enquote {\bibinfo {title} {Subdiffraction
  incoherent optical imaging via spatial-mode demultiplexing: {Semiclassical}
  treatment},}\ }\href {\doibase 10.1103/PhysRevA.97.023830} {\bibfield
  {journal} {\bibinfo  {journal} {Physical Review A}\ }\textbf {\bibinfo
  {volume} {97}},\ \bibinfo {pages} {023830} (\bibinfo {year}
  {2018})}\BibitemShut {NoStop}%
\bibitem [{\citenamefont {Tsang}(2019{\natexlab{c}})}]{spade_prr}%
  \BibitemOpen
  \bibfield  {author} {\bibinfo {author} {\bibfnamefont {Mankei}\ \bibnamefont
  {Tsang}},\ }\bibfield  {title} {\enquote {\bibinfo {title} {Semiparametric
  estimation for incoherent optical imaging},}\ }\href {\doibase
  10.1103/PhysRevResearch.1.033006} {\bibfield  {journal} {\bibinfo  {journal}
  {Physical Review Research}\ }\textbf {\bibinfo {volume} {1}},\ \bibinfo
  {pages} {033006} (\bibinfo {year} {2019}{\natexlab{c}})}\BibitemShut
  {NoStop}%
\bibitem [{\citenamefont {Tsang}(2023{\natexlab{a}})}]{superosc_ieee}%
  \BibitemOpen
  \bibfield  {author} {\bibinfo {author} {\bibfnamefont {Mankei}\ \bibnamefont
  {Tsang}},\ }\bibfield  {title} {\enquote {\bibinfo {title} {Efficient
  superoscillation measurement for incoherent optical imaging},}\ }\href
  {\doibase 10.1109/JSTSP.2022.3212173} {\bibfield  {journal} {\bibinfo
  {journal} {IEEE Journal of Selected Topics in Signal Processing}\ }\textbf
  {\bibinfo {volume} {17}},\ \bibinfo {pages} {513–524} (\bibinfo {year}
  {2023}{\natexlab{a}})}\BibitemShut {NoStop}%
\bibitem [{\citenamefont {Tsang}(2023{\natexlab{b}})}]{noise_spec_pra}%
  \BibitemOpen
  \bibfield  {author} {\bibinfo {author} {\bibfnamefont {Mankei}\ \bibnamefont
  {Tsang}},\ }\bibfield  {title} {\enquote {\bibinfo {title} {Quantum noise
  spectroscopy as an incoherent imaging problem},}\ }\href {\doibase
  10.1103/PhysRevA.107.012611} {\bibfield  {journal} {\bibinfo  {journal}
  {Physical Review A}\ }\textbf {\bibinfo {volume} {107}},\ \bibinfo {pages}
  {012611} (\bibinfo {year} {2023}{\natexlab{b}})}\BibitemShut {NoStop}%
\bibitem [{\citenamefont {Bisketzi}\ \emph {et~al.}(2019)\citenamefont
  {Bisketzi}, \citenamefont {Branford},\ and\ \citenamefont
  {Datta}}]{bisketzi19}%
  \BibitemOpen
  \bibfield  {author} {\bibinfo {author} {\bibfnamefont {Evangelia}\
  \bibnamefont {Bisketzi}}, \bibinfo {author} {\bibfnamefont {Dominic}\
  \bibnamefont {Branford}}, \ and\ \bibinfo {author} {\bibfnamefont {Animesh}\
  \bibnamefont {Datta}},\ }\bibfield  {title} {\enquote {\bibinfo {title}
  {Quantum limits of localisation microscopy},}\ }\href {\doibase
  10.1088/1367-2630/ab58a0} {\bibfield  {journal} {\bibinfo  {journal} {New
  Journal of Physics}\ }\textbf {\bibinfo {volume} {21}},\ \bibinfo {pages}
  {123032} (\bibinfo {year} {2019})}\BibitemShut {NoStop}%
\bibitem [{\citenamefont {Dutton}\ \emph {et~al.}(2019)\citenamefont {Dutton},
  \citenamefont {Kerviche}, \citenamefont {Ashok},\ and\ \citenamefont
  {Guha}}]{dutton19}%
  \BibitemOpen
  \bibfield  {author} {\bibinfo {author} {\bibfnamefont {Zachary}\ \bibnamefont
  {Dutton}}, \bibinfo {author} {\bibfnamefont {Ronan}\ \bibnamefont
  {Kerviche}}, \bibinfo {author} {\bibfnamefont {Amit}\ \bibnamefont {Ashok}},
  \ and\ \bibinfo {author} {\bibfnamefont {Saikat}\ \bibnamefont {Guha}},\
  }\bibfield  {title} {\enquote {\bibinfo {title} {Attaining the quantum limit
  of superresolution in imaging an object's length via predetection
  spatial-mode sorting},}\ }\href {\doibase 10.1103/PhysRevA.99.033847}
  {\bibfield  {journal} {\bibinfo  {journal} {Physical Review A}\ }\textbf
  {\bibinfo {volume} {99}},\ \bibinfo {pages} {033847} (\bibinfo {year}
  {2019})}\BibitemShut {NoStop}%
\bibitem [{\citenamefont {Prasad}(2020)}]{prasad20}%
  \BibitemOpen
  \bibfield  {author} {\bibinfo {author} {\bibfnamefont {Sudhakar}\
  \bibnamefont {Prasad}},\ }\bibfield  {title} {\enquote {\bibinfo {title}
  {Quantum limited superresolution of extended sources in one and two
  dimensions},}\ }\href {\doibase 10.1103/PhysRevA.102.063719} {\bibfield
  {journal} {\bibinfo  {journal} {Physical Review A}\ }\textbf {\bibinfo
  {volume} {102}},\ \bibinfo {pages} {063719} (\bibinfo {year}
  {2020})}\BibitemShut {NoStop}%
\bibitem [{\citenamefont {Prasad}(2023)}]{prasad23}%
  \BibitemOpen
  \bibfield  {author} {\bibinfo {author} {\bibfnamefont {Sudhakar}\
  \bibnamefont {Prasad}},\ }\bibfield  {title} {\enquote {\bibinfo {title}
  {Quantum limits on localizing point objects against a uniformly bright
  disk},}\ }\href {\doibase 10.1103/PhysRevA.107.032427} {\bibfield  {journal}
  {\bibinfo  {journal} {Physical Review A}\ }\textbf {\bibinfo {volume}
  {107}},\ \bibinfo {pages} {032427} (\bibinfo {year} {2023})}\BibitemShut
  {NoStop}%
\bibitem [{\citenamefont {Tsang}(2011)}]{stellar}%
  \BibitemOpen
  \bibfield  {author} {\bibinfo {author} {\bibfnamefont {Mankei}\ \bibnamefont
  {Tsang}},\ }\bibfield  {title} {\enquote {\bibinfo {title} {Quantum
  nonlocality in weak-thermal-light interferometry},}\ }\href {\doibase
  10.1103/PhysRevLett.107.270402} {\bibfield  {journal} {\bibinfo  {journal}
  {Physical Review Letters}\ }\textbf {\bibinfo {volume} {107}},\ \bibinfo
  {pages} {270402} (\bibinfo {year} {2011})}\BibitemShut {NoStop}%
\bibitem [{\citenamefont {Tsang}(2021{\natexlab{b}})}]{poisson_quantum}%
  \BibitemOpen
  \bibfield  {author} {\bibinfo {author} {\bibfnamefont {Mankei}\ \bibnamefont
  {Tsang}},\ }\bibfield  {title} {\enquote {\bibinfo {title} {Poisson {Quantum}
  {Information}},}\ }\href {\doibase 10.22331/q-2021-08-19-527} {\bibfield
  {journal} {\bibinfo  {journal} {Quantum}\ }\textbf {\bibinfo {volume} {5}},\
  \bibinfo {pages} {527} (\bibinfo {year} {2021}{\natexlab{b}})}\BibitemShut
  {NoStop}%
\bibitem [{\citenamefont {Gross}\ and\ \citenamefont {Caves}(2020)}]{gross20}%
  \BibitemOpen
  \bibfield  {author} {\bibinfo {author} {\bibfnamefont {Jonathan~Arthur}\
  \bibnamefont {Gross}}\ and\ \bibinfo {author} {\bibfnamefont {Carlton~M.}\
  \bibnamefont {Caves}},\ }\bibfield  {title} {\enquote {\bibinfo {title} {One
  from many: {Estimating} a function of many parameters},}\ }\href {\doibase
  10.1088/1751-8121/abb9ed} {\bibfield  {journal} {\bibinfo  {journal} {Journal
  of Physics A: Mathematical and Theoretical}\ }\textbf {\bibinfo {volume}
  {54}},\ \bibinfo {pages} {014001} (\bibinfo {year} {2020})}\BibitemShut
  {NoStop}%
\bibitem [{\citenamefont {Hayashi}(2017)}]{hayashi}%
  \BibitemOpen
  \bibfield  {author} {\bibinfo {author} {\bibfnamefont {Masahito}\
  \bibnamefont {Hayashi}},\ }\href {\doibase 10.1007/978-3-662-49725-8} {\emph
  {\bibinfo {title} {Quantum {I}nformation {T}heory: {M}athematical
  {F}oundation}}},\ \bibinfo {edition} {2nd}\ ed.\ (\bibinfo  {publisher}
  {Springer},\ \bibinfo {address} {Berlin},\ \bibinfo {year}
  {2017})\BibitemShut {NoStop}%
\bibitem [{\citenamefont {Tsang}(2022)}]{gce_pra}%
  \BibitemOpen
  \bibfield  {author} {\bibinfo {author} {\bibfnamefont {Mankei}\ \bibnamefont
  {Tsang}},\ }\bibfield  {title} {\enquote {\bibinfo {title} {Generalized
  conditional expectations for quantum retrodiction and smoothing},}\ }\href
  {\doibase 10.1103/PhysRevA.105.042213} {\bibfield  {journal} {\bibinfo
  {journal} {Physical Review A}\ }\textbf {\bibinfo {volume} {105}},\ \bibinfo
  {pages} {042213} (\bibinfo {year} {2022})}\BibitemShut {NoStop}%
\bibitem [{\citenamefont {Tsang}(2023{\natexlab{c}})}]{gce2}%
  \BibitemOpen
  \bibfield  {author} {\bibinfo {author} {\bibfnamefont {Mankei}\ \bibnamefont
  {Tsang}},\ }\bibfield  {title} {\enquote {\bibinfo {title} {Operational
  meanings of a generalized conditional expectation in quantum metrology},}\
  }\href {\doibase 10.22331/q-2023-11-03-1162} {\bibfield  {journal} {\bibinfo
  {journal} {Quantum}\ }\textbf {\bibinfo {volume} {7}},\ \bibinfo {pages}
  {1162} (\bibinfo {year} {2023}{\natexlab{c}})},\ \Eprint
  {http://arxiv.org/abs/2212.13162v6} {2212.13162v6} \BibitemShut {NoStop}%
\bibitem [{\citenamefont {Řeháček}\ \emph {et~al.}(2017)\citenamefont
  {Řeháček}, \citenamefont {Paúr}, \citenamefont {Stoklasa}, \citenamefont
  {Hradil},\ and\ \citenamefont {Sánchez-Soto}}]{rehacek17}%
  \BibitemOpen
  \bibfield  {author} {\bibinfo {author} {\bibfnamefont {J.}~\bibnamefont
  {Řeháček}}, \bibinfo {author} {\bibfnamefont {M.}~\bibnamefont {Paúr}},
  \bibinfo {author} {\bibfnamefont {B.}~\bibnamefont {Stoklasa}}, \bibinfo
  {author} {\bibfnamefont {Z.}~\bibnamefont {Hradil}}, \ and\ \bibinfo {author}
  {\bibfnamefont {L.~L.}\ \bibnamefont {Sánchez-Soto}},\ }\bibfield  {title}
  {\enquote {\bibinfo {title} {Optimal measurements for resolution beyond the
  {Rayleigh} limit},}\ }\href {\doibase 10.1364/OL.42.000231} {\bibfield
  {journal} {\bibinfo  {journal} {Optics Letters}\ }\textbf {\bibinfo {volume}
  {42}},\ \bibinfo {pages} {231–234} (\bibinfo {year} {2017})}\BibitemShut
  {NoStop}%
\bibitem [{\citenamefont {Olver}\ \emph {et~al.}(2010)\citenamefont {Olver},
  \citenamefont {Lozier}, \citenamefont {Boisvert},\ and\ \citenamefont
  {Clark}}]{olver}%
  \BibitemOpen
  \bibinfo {editor} {\bibfnamefont {F.~W.~J.}\ \bibnamefont {Olver}}, \bibinfo
  {editor} {\bibfnamefont {D.~W.}\ \bibnamefont {Lozier}}, \bibinfo {editor}
  {\bibfnamefont {R.~F.}\ \bibnamefont {Boisvert}}, \ and\ \bibinfo {editor}
  {\bibfnamefont {C.~W.}\ \bibnamefont {Clark}},\ eds.,\ \href@noop {} {\emph
  {\bibinfo {title} {NIST Handbook of Mathematical Functions}}}\ (\bibinfo
  {publisher} {NIST and Cambridge University Press},\ \bibinfo {address}
  {Cambridge},\ \bibinfo {year} {2010})\BibitemShut {NoStop}%
\bibitem [{\citenamefont {Holevo}(2011)}]{holevo11}%
  \BibitemOpen
  \bibfield  {author} {\bibinfo {author} {\bibfnamefont {Alexander~S.}\
  \bibnamefont {Holevo}},\ }\href {\doibase 10.1007/978-88-7642-378-9} {\emph
  {\bibinfo {title} {Probabilistic and Statistical Aspects of Quantum
  Theory}}}\ (\bibinfo  {publisher} {Scuola Normale Superiore Pisa},\ \bibinfo
  {address} {Pisa, Italy},\ \bibinfo {year} {2011})\BibitemShut {NoStop}%
\bibitem [{\citenamefont {Carollo}\ \emph {et~al.}(2019)\citenamefont
  {Carollo}, \citenamefont {Spagnolo}, \citenamefont {Dubkov},\ and\
  \citenamefont {Valenti}}]{carollo19}%
  \BibitemOpen
  \bibfield  {author} {\bibinfo {author} {\bibfnamefont {Angelo}\ \bibnamefont
  {Carollo}}, \bibinfo {author} {\bibfnamefont {Bernardo}\ \bibnamefont
  {Spagnolo}}, \bibinfo {author} {\bibfnamefont {Alexander~A.}\ \bibnamefont
  {Dubkov}}, \ and\ \bibinfo {author} {\bibfnamefont {Davide}\ \bibnamefont
  {Valenti}},\ }\bibfield  {title} {\enquote {\bibinfo {title} {On quantumness
  in multi-parameter quantum estimation},}\ }\href {\doibase
  10.1088/1742-5468/ab3ccb} {\bibfield  {journal} {\bibinfo  {journal} {Journal
  of Statistical Mechanics: Theory and Experiment}\ }\textbf {\bibinfo {volume}
  {2019}},\ \bibinfo {pages} {094010} (\bibinfo {year} {2019})}\BibitemShut
  {NoStop}%
\bibitem [{\citenamefont {Carollo}\ \emph {et~al.}(2020)\citenamefont
  {Carollo}, \citenamefont {Spagnolo}, \citenamefont {Dubkov},\ and\
  \citenamefont {Valenti}}]{carollo20}%
  \BibitemOpen
  \bibfield  {author} {\bibinfo {author} {\bibfnamefont {Angelo}\ \bibnamefont
  {Carollo}}, \bibinfo {author} {\bibfnamefont {Bernardo}\ \bibnamefont
  {Spagnolo}}, \bibinfo {author} {\bibfnamefont {Alexander~A.}\ \bibnamefont
  {Dubkov}}, \ and\ \bibinfo {author} {\bibfnamefont {Davide}\ \bibnamefont
  {Valenti}},\ }\bibfield  {title} {\enquote {\bibinfo {title} {Erratum: {On}
  quantumness in multi-parameter quantum estimation (2019 {J}. {Stat}. {Mech}.
  094010)},}\ }\href {\doibase 10.1088/1742-5468/ab6f5e} {\bibfield  {journal}
  {\bibinfo  {journal} {Journal of Statistical Mechanics: Theory and
  Experiment}\ }\textbf {\bibinfo {volume} {2020}},\ \bibinfo {pages} {029902}
  (\bibinfo {year} {2020})}\BibitemShut {NoStop}%
\bibitem [{\citenamefont {Nagaoka}(1989)}]{nagaoka89}%
  \BibitemOpen
  \bibfield  {author} {\bibinfo {author} {\bibfnamefont {Hiroshi}\ \bibnamefont
  {Nagaoka}},\ }\bibfield  {title} {\enquote {\bibinfo {title} {A new approach
  to cramér-rao bounds for quantum state estimation},}\ }\href@noop {}
  {\bibfield  {journal} {\bibinfo  {journal} {IEICE Technical Report}\ }\textbf
  {\bibinfo {volume} {IT 89-42}},\ \bibinfo {pages} {9–14} (\bibinfo {year}
  {1989})}\BibitemShut {NoStop}%
\bibitem [{\citenamefont {Hayashi}(1999)}]{hayashi99}%
  \BibitemOpen
  \bibfield  {author} {\bibinfo {author} {\bibfnamefont {Masahito}\
  \bibnamefont {Hayashi}},\ }\bibfield  {title} {\enquote {\bibinfo {title} {On
  simultaneous measurement of noncommutative observables},}\ }in\ \href
  {https://www.kurims.kyoto-u.ac.jp/~kyodo/kokyuroku/contents/pdf/1099-8.pdf}
  {\emph {\bibinfo {booktitle} {Development of Infinite-Dimensional
  Noncommutative Analysis}}},\ \bibinfo {series and number} {\bibinfo {series}
  {RIMS Kokyuroku}\ No.\ \bibinfo {number} {1099}},\ \bibinfo {editor} {edited
  by\ \bibinfo {editor} {\bibfnamefont {Akihito}\ \bibnamefont {Hora}}}\
  (\bibinfo  {publisher} {Kyoto University},\ \bibinfo {address} {Kyoto},\
  \bibinfo {year} {1999})\ p.\ \bibinfo {pages} {96–118},\ \bibinfo {note}
  {in Japanese}\BibitemShut {NoStop}%
\bibitem [{\citenamefont {Conlon}\ \emph {et~al.}(2021)\citenamefont {Conlon},
  \citenamefont {Suzuki}, \citenamefont {Lam},\ and\ \citenamefont
  {Assad}}]{conlon21}%
  \BibitemOpen
  \bibfield  {author} {\bibinfo {author} {\bibfnamefont
  {Lorc{\ifmmode\acute{a}\elseá\fi}n~O.}\ \bibnamefont {Conlon}}, \bibinfo
  {author} {\bibfnamefont {Jun}\ \bibnamefont {Suzuki}}, \bibinfo {author}
  {\bibfnamefont {Ping~Koy}\ \bibnamefont {Lam}}, \ and\ \bibinfo {author}
  {\bibfnamefont {Syed~M.}\ \bibnamefont {Assad}},\ }\bibfield  {title}
  {\enquote {\bibinfo {title} {Efficient computation of the
  {N}agaoka–{H}ayashi bound for multiparameter estimation with separable
  measurements},}\ }\href {\doibase 10.1038/s41534-021-00414-1} {\bibfield
  {journal} {\bibinfo  {journal} {npj Quantum Information}\ }\textbf {\bibinfo
  {volume} {7}},\ \bibinfo {pages} {1–8} (\bibinfo {year}
  {2021})}\BibitemShut {NoStop}%
\end{thebibliography}%

\end{document}